\documentclass[a4paper,11pt]{article}
\usepackage{pos}
\usepackage{comment}
\usepackage[symbol]{footmisc}
\renewcommand{\thefootnote}{\fnsymbol{footnote}}

\title{Results on Low-Mass Weakly Interacting Massive Particles from a 11 kg d Target Exposure of DAMIC at SNOLAB}
 \ShortTitle{WIMP results from 11 kg d target exposure of DAMIC at SNOLAB}

\author*[a]{Michelangelo Traina}

\affiliation[a]{Laboratoire de Physique Nucléaire et des Hautes Energies (LPNHE),\\
  Sorbonne Université, CNRS-IN2P3, Paris, France}

\forColl{DAMIC} 

\emailAdd{michelangelo.traina@lpnhe.in2p3.fr}

\abstract{Experimental efforts of the last decades have been unsuccessful in detecting WIMPs (Weakly Interacting Massive Particles) in the 10-to-10$^4$~GeV/$c^2$ range, thus motivating the search for lighter dark matter. The DAMIC (DArk Matter In CCDs) at SNOLAB experiment aims for direct detection of light dark matter particles ($m_\chi<10$~GeV/$c^2$) by means of CCDs (Charge-Coupled Devices). Fully-depleted 675~$\mu$m-thick CCDs are used to such end. The optimized readout noise and operation at cryogenic temperatures allow for a detection threshold of 50~eV$_\text{ee}$ electron-equivalent energy. Focusing on nuclear and electronic scattering as potential detection processes, DAMIC has so far set competitive constraints on the detection of low mass WIMPs and hidden-sector particles.
In this work, an 11~kg$\cdot$d exposure dataset is exploited to search for light WIMPs by building the first comprehensive radioactive background model for CCDs. Different background sources are discriminated making conjoint use of the spatial distribution and energy of ionization events, thereby constraining the amount of contaminants such as tritium from silicon cosmogenic activation and surface lead-210 from radon plate-out.
Despite a conspicuous, statistically-significant excess of events below 200~eV$_\text{ee}$, this analysis places the strongest exclusion limit on the WIMP-nucleon scattering cross section with a silicon target for 1~GeV<$m_\chi c^2$~<~9~GeV.}

\FullConference{37$^{\rm{th}}$ International Cosmic Ray Conference (ICRC 2021)\\
		July 12th -- 23rd, 2021\\
		Online -- Berlin, Germany}

\begin{document}
\maketitle

\section{Introduction}
There is abundant astronomical and cosmological evidence for the existence of cold, non-baryonic dark matter (DM). Galactic rotation curves and the Cosmic Microwave Background (CMB) acoustic peaks are among the most important and best known pieces of such evidence, but do not constitute its entirety. Thermal Weakly Interacting Massive Particles (WIMPs) are a popular cold dark matter candidate and have been searched for during the last decades, unsuccessfully. The main reason behind the interest on WIMPs is the so-called \textit{WIMP miracle}. This is an outstanding coincidence, in that plugging in weak-scale numbers for the WIMP pair-annihilation cross section ($\sim10^{-36}$~cm$^2$) and WIMP mass provides the correct observed value for the dark matter relic density \cite{appec}. As the community set ever stronger limits on the existence of conventional WIMPs ($m_\chi~\gtrsim~10~$GeV/$c^2$), interest partly shifted to lighter dark matter, among which hidden-sector candidates, involving Standard Model-Dark Matter interactions mediated by mixed messenger states.

The DAMIC experiments (Dark Matter in CCDs) at SNOLAB and, in the near future, at Modane, France \cite{damic,damicm}, look to detecting light WIMPs ($1$ GeV~$ \lesssim m_\chi c^2~\lesssim~10$~GeV) and/or hidden-sector candidates by means of charge-coupled devices (CCDs). Silicon devices characterized by a three-phase polysilicon gate structure, CCDs collect ionization-produced charge carriers in a p-doped buried channel less than 1~$\mu$m below the gates, and then transfer it to a read-out amplifier where they are converted in a voltage signal. The ionizing particles can be recoiled nuclei or electrons, depending on the DM candidate assumption, with a target sensitive to energy depositions as low as 1.1~eV. This allows to set limits on the cross sections of light WIMPs, hidden-sector particles and their massive mediators. The DAMIC collaboration has achieved several such results since its birth in 2011 (\textit{e.g.}, see \cite{ws2016,hsdm2019}). This article provides an overview of the latest 11~kg~day WIMP search of DAMIC at SNOLAB, overall taking place between 2017 and 2020 \cite{ws2020}.

\section{Experimental Setup}

The DAMIC experiment is located at the SNOLAB underground laboratory in Sudbury, Ontario, Canada, beneath a 2 km granite overburden (6 km water equivalent) \cite{snolab}. The salient part of the DAMIC detector are its sensors: high-resistivity ($\sim$10~k$\Omega \cdot$cm), fully depleted CCDs. DAMIC CCDs are 674$\pm$3~$\mu$m thick with an active thickness of 665$\pm$5~$\mu$m. Each device weighs 6.0~g and consists of an array of 4116$\times$4128 pixels of 15~$\mu$m $\times$ 15~$\mu$m area. The sensor active region is fully depleted by means of a bias voltage of 70~V applied to a backside planar contact. When charge is produced by interactions in the active region, the positive carriers drift in the direction of the electric field, until they are collected in the buried-channel potential wells. The substrate voltage determines the lateral diffusion that carriers undergo as they drift towards the gates, producing the observed lateral spread of tracks on the CCD $xy$ plane. A measure of the latter can be used to reconstruct the interaction point in the sensor in 3D, in virtue of its proportionality with the carrier transit time. 
At the end of the exposure period, charges are transferred for read-out by means of parallel and serial clock voltages, corresponding to row- and column-wise transfer, respectively. As the serial register is read out, successive rows of pixel charge are therein transferred by the parallel clocks until all rows are read out. 

\begin{figure}[t]
	\centering
	\includegraphics[width=0.47\textwidth]{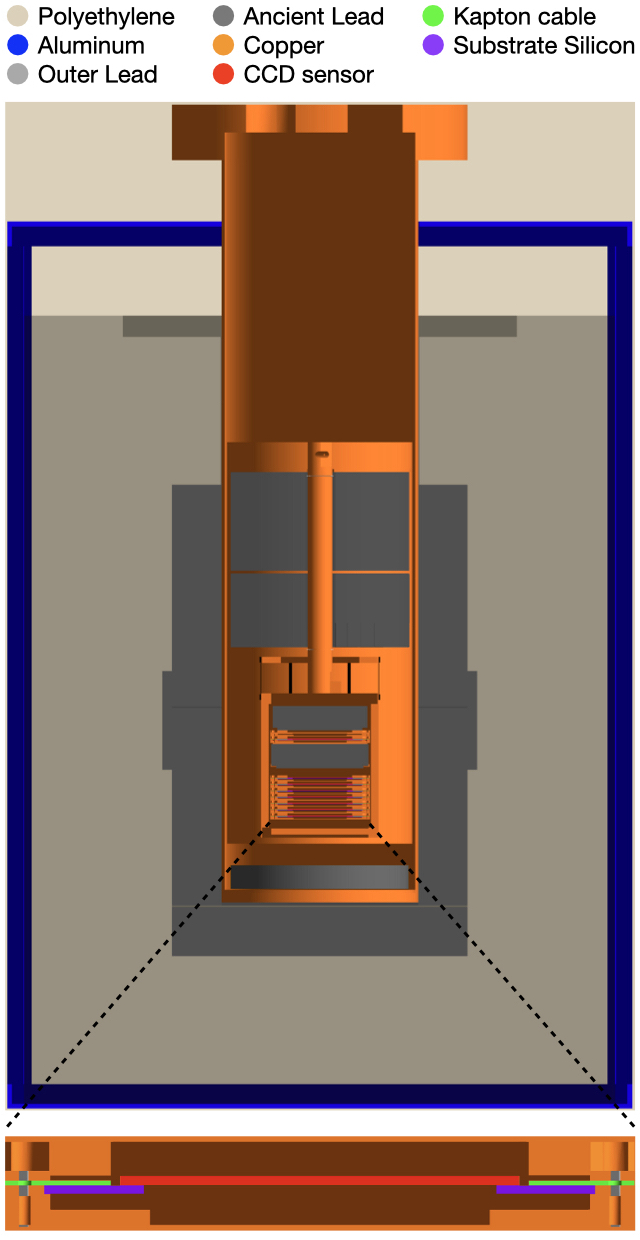}
	\caption{Cross-section of the GEANT4 geometry of the DAMIC at SNOLAB detector \cite{bkgpaper}, with same-material detector parts indicated by color according to the top legend. Only the upper polyethylene layer is shown. A zoomed modified aspect ratio view of a CCD module is reported in the bottom of the image.}
	\label{fig:geom}
\end{figure}

DAMIC at SNOLAB CCDs are protected from cosmic radiation and external radioactivity by means of multiple layers of shielding. Each CCD is packaged into a OFHC (oxygen-free high thermal conductivity) copper module containing the sensor, the kapton flex readout cable and the silicon frame they are glued onto. The 8-CCD tower is installed into a 6.35~mm thick OFHC copper box serving as a cold IR shielding. The box is located in between 2.5~cm (bottom) and 18~cm (top) of lead shielding, inside a cylindrical copper cryostat. 
The cryostat is then placed inside a lead shield, of which the innermost 5~cm is archaeological \cite{spanishLead}, and the outer 15~cm is low activity \cite{exo}. Outside of this lead shield, a 42~cm polyethylene layer provides protection against external neutrons. The system is held at high vacuum ($10^{-6}$~mbar) and cryogenic temperature of 140~K. Boil-off nitrogen from an LN$_2$ dewar with a flow rate of 2~L/min is used to purge the volume inside the shielding from radon. A simulated cross-sectional geometry of the DAMIC at SNOLAB detector is shown in Figure \ref{fig:geom}.

\section{Backgrounds}
The sensitivity of dark matter direct detection experiments is ultimately determined by the rate of radioactive backgrounds that mimic the signal of the relevant candidates. 
DAMIC addresses such backgrounds through mitigation and rejection. 

Mitigation relies on underground operation, material selection and in situ shielding. Cosmic radiation flux is suppressed by a factor $\sim5\cdot10^7$ going from sea level to underground operation at SNOLAB \cite{snolab}. \begin{figure}[t]
	\centering
	\includegraphics[width=0.6\textwidth]{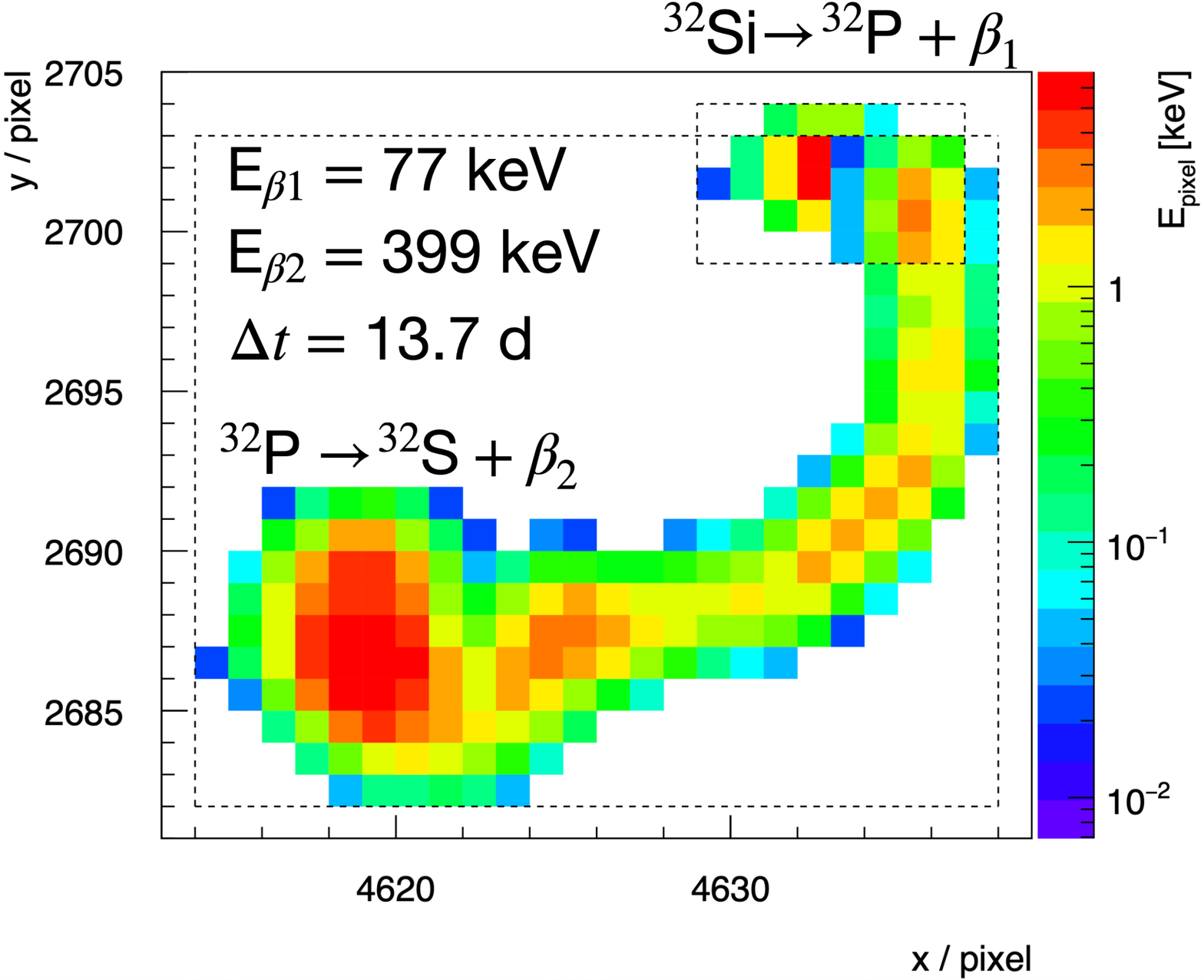}
	\caption{$^{32}$Si spatial decay coincidence candidate \cite{bulkcontamination}. The upper cluster within the smaller dashed rectangle is associated to a $^{32}$Si $\beta$ decay into $^{32}$P, and has an energy of 77~keV. The lower cluster contained in the bigger rectangle occurred 13.7~days later and is associated to a $^{32}$P $\beta$ decay into $^{32}$S, having an energy of 399~keV.}
	\label{fig:si32}
\vspace{0pt}
\end{figure} Detector materials are selected and assayed to ensure high radiopurity. In particular, low levels of long-lived primordial isotopes such as $^{238}$U and $^{232}$Th. Most materials in use in the DAMIC at SNOLAB detector were assayed using Inductively-Coupled Mass Spectrometry (ICP-MS), Glow Discharge Mass Spectrometry (GDMS) or Germanium $\gamma$-ray spectrometry at the SNOLAB $\gamma$-ray counting facility \cite{snolab}. Activity values for main contaminants are reported in Table \ref{tab:parttable}, for detector parts relevant for the background model construction.

Rejection involves the discrimination of decay events from potential dark matter signals, and the subsequent quantification (or upper-limit setting) of the corresponding radioactive contaminants: precious information for one's background model. Advantageously, CCDs can be used to identify and quantify many of the contaminants they carry through a spatial coincidence analysis. In practice, in-sensor radioisotopes at the top of a decay chain are measured by studying the properties of spatially coinciding events, selecting those whose energy, spatial distribution and time separation fall in selection intervals optimized for different decays \cite{bulkcontamination}. An example of $^{32}$Si spatial coincidence candidate is reported in Figure \ref{fig:si32}.
This capability can guide material selection – \textit{e.g.}, by measuring $^{32}$Si traces in the bulk of the sensor. Here, it is used to enhance rejection by providing constraints used in the construction of our background model.
Main CCD contaminants intrinsic $^{32}$Si and $^{210}$Pb from radon plate-out have been quantified in this fashion. Measurements of their activities stemming from this analysis are reported in Table \ref{tab:parttable}.

Neutrons and residual cosmic rays penetrating the SNOLAB cavern are also rejected. Extensive GEANT4 \cite{geant4} simulations have shown that ionization events produced by neutrons and neutron-induced radiation are negligible, being orders of magnitude below the observed event rate \cite{bkgpaper}. The cosmic muon flux at SNOLAB of 0.27~m$^{-2}$ d$^{-1}$ \cite{snolab} is equivalent to 1~muon crossing the CCD tower every $\sim$10$^3$~days. Such muon would have mean kinetic energy >~300~GeV \cite{meiandhime}, triggering an important particle shower in the detector, with numerous clusters coinciding in multiple CCDs during the same exposure. Similar events were not detected in the WIMP search run.

\begin{table*}[!h]
\begin{center}
\resizebox{\textwidth}{!}{
\begin{tabular}{r|@{\hskip 0.1in} r@{\hskip 0.15in} r@{\hskip 0.15in} r@{\hskip 0.15in} r@{\hskip 0.15in} r@{\hskip 0.15in} r}
\hline \hline
\rule{0pt}{2.5ex} & \multicolumn{1}{c}{$^{238}$U} & \multicolumn{1}{c}{$^{226}$Ra} & \multicolumn{1}{c}{$^{210}$Pb} & \multicolumn{1}{c}{$^{232}$Th} & \multicolumn{1}{c}{$^{40}$K} & \multicolumn{1}{c}{$^{32}$Si}\\
\hline
\rule{0pt}{2.5ex}CCD / Si frame &$<$11~{\cite{bulkcontamination}} &$<$5.3~{\cite{bulkcontamination}} & $<$160*~{\cite{bulkcontamination}} &$<$7.3~{\cite{bulkcontamination}} &$<$0.5 $^{[M]}$ &$140 \pm 30$~{\cite{bulkcontamination}} \\
 Kapton Cable & 58000 $\pm$ 5000 $^{[M]}$ & 4900 $\pm$ 5700 $^{[G]}$ & ...
& 3200 $\pm$ 500 $^{[M]}$ & 29000 $\pm$ 2000 $^{[M]}$ & ...\\
OFHC Copper & $<$120 $^{[M]}$ &$<$130 $^{[G]}$ & 27000 $\pm$ 8000~\cite{xmass}
&$<$41 $^{[M]}$ &$<$31 $^{[M]}$ & ...\\
Module Screws & 16000 $\pm$ 44000 $^{[G]}$ &$<$138 $^{[G]}$ &  27000 $\pm$ 8000$^\dagger$ 
& 2300 $\pm$ 1600 $^{[G]}$ & 28000 $\pm$ 15000 $^{[G]}$ & ...\\
Ancient Lead &$<$23~\cite{spanishLead} & $<$260 $^{[G]}$ & 33000$^\ddagger$\cite{spanishLead}
& 2.3$^\ddagger$\cite{spanishLead} &$<$5.8 $^{[M]}$ & ...\\
Outer Lead &$<$13~{\cite{exo}} &  $<$200 $^{[G]}$ & ($19\pm5$)$\times 10^6$~\cite{exo} &$<$4.6~{\cite{exo}} &$<$220~{\cite{exo}} & ...\\
\hline \hline
\end{tabular}}
\caption{\label{tab:parttable}Activities used to constrain the amount of radioactivity in each simulated detector part, in units of $\mu$Bq/kg. To be published in \cite{bkgpaper}. A superscript $M$ ($G$) is used for mass spectrometry (gamma counting) assay values. An asterisk* indicates a template which is not used in the background model analysis (but listed for completeness). A dagger$^\dagger$  indicates a preliminary assumption which has minimal ($<$0.1 dru) effect on the final background model. A double-dagger$^\ddagger$  indicates a measurement which is fixed (given 1$\%$ uncertainty) in the background model analysis to reduce degeneracy among the fit parameters.}
\end{center}
\end{table*}

\subsection{Background Model Construction}

To simulate the spectra of ionization events from main radioactive backgrounds, we utilize the GEANT4 simulation package. 
Simulations are executed in detector parts grouped by common material origin, as summarized in Table \ref{tab:parttable}. 
Relevant isotopes come from the $^{238}$U, $^{232}$Th and $^{40}$K chains, or material-specific isotopes (\textit{e.g.}, from activation). Major contributors are $^3$H, $^{22}$Na, $^{32}$Si and $^{210}$Pb. An important fraction of these has recent history in the detector materials, originating during production, storage or transportation. This shows that improved protocols have the potential to considerably reduce the background rates affecting DAMIC detectors.

GEANT4 raw data is reconstructed to translate it into a CCD image, implementing the detector response (\textit{e.g}, readout noise, digital resolution, pixel binning\footnote{Rows and/or columns of pixel charge cumulated in a single pixel. Binning can improve signal to noise ratio.}). Data is then clustered, compressing it to a list of cluster variables, among which the energy, E, and the lateral spread, $\sigma_x$. 

\begin{figure}[t!]
	\centering
	\includegraphics[width=0.99\textwidth]{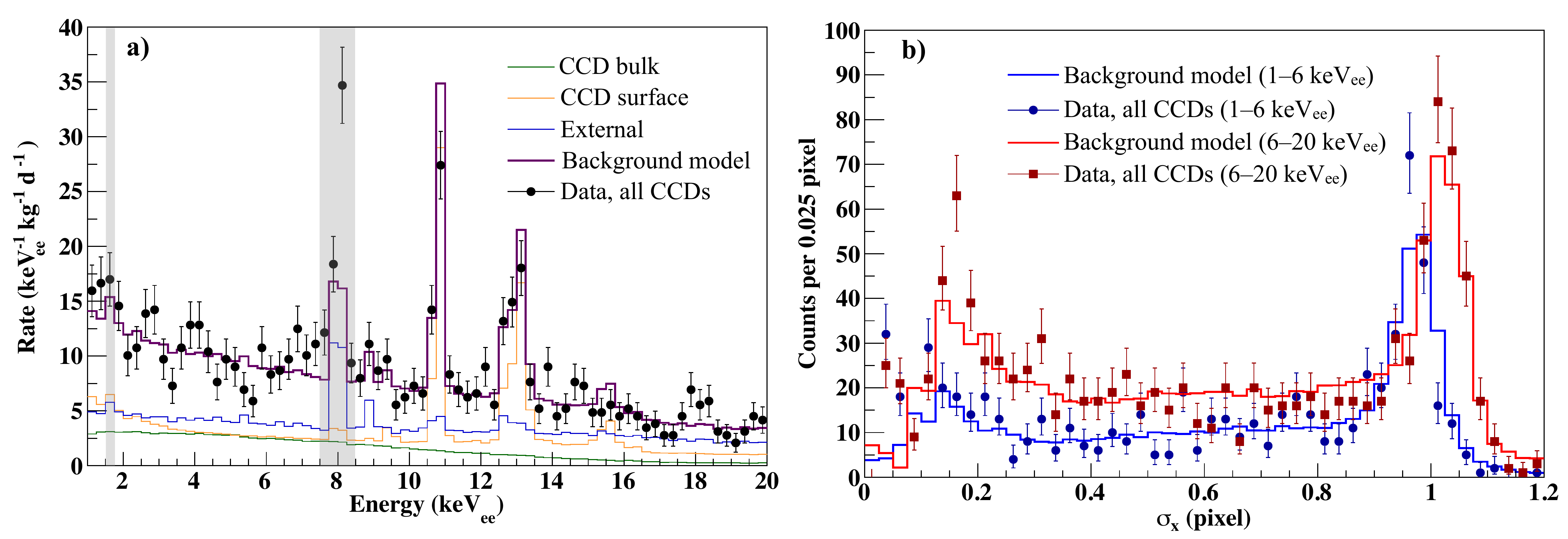}
	\caption{Projections in E and $\sigma_x$ of the best-fit background model (solid lines) compared to the data (markers) \cite{ws2020}. a) The total background model spectrum is shown along with individual contributions from different background sources both in the fit ($6-20$ keV$_\text{ee}$) and prediction range ($0-6$ keV$_\text{ee}$). Shaded energy regions are excluded from the analysis. b) Comparison of $\sigma_x$ distributions in the fit energy region (red) and extrapolated toward lower energies (blue). The peak at low (high) $\sigma_x$ corresponds to events at the front (back) of the CCDs.}
	\label{fig:bkg}
\end{figure}

The background model is constructed by performing a two-dimensional binned-likelihood fit to data. 
Templates sorted by detector part and decay chain are constructed in the 6-20~keV$_\text{ee}$ range, where no WIMP signal is expected\footnote{Experiments have placed strong limits on WIMPs  $m_\chi>10$ GeV/$c^2$ whose recoil signals populate this energy range.}, while covering the full tritium beta spectrum (a dominant bulk contaminant). 
Only data from CCD 2-7 is used for the fit, as CCD 1 features a different background environment, being sandwiched between two ancient lead bricks (see Figure \ref{fig:geom}). 
The two-dimensional log-likelihood is calculated as the sum over all energy and sigma bins assuming the bin content follows a Poisson distribution, with additional penalty terms to constrain the model according to assay results.
We  fit  the two-dimensional E-$\sigma_x$ distribution for  CCDs 2-7  and compare the result against CCD 1 as a check. Fit result projections for all CCDs are shown in Figure \ref{fig:bkg}.

\section{WIMP Search}
\begin{figure}[t]
	\centering
	\includegraphics[width=0.99\textwidth]{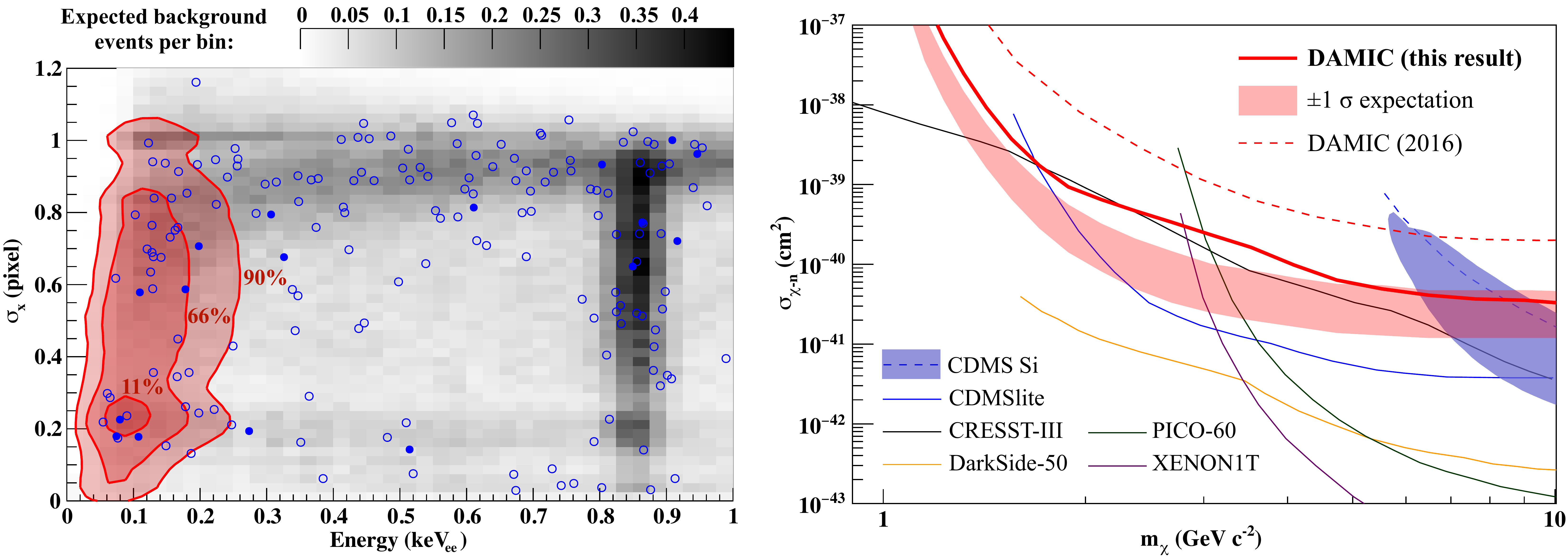}
	\caption{
\textbf{Left:} Data, background model and excess in (E,$\sigma_x$) space \cite{ws2020}. Blue data points are overlaid on the background model in gray. Open circles correspond to CCDs 2–7, filled circles correspond to CCD 1. Red contours represent the best-fit exponential excess. The Ne K de-excitation line (0.85 keV$_\text{ee}$) emitted post electron capture by $^{22}$Na in the CCD bulk is visible. \textbf{Right:} Upper limit (90\% CL) on $\sigma_{\chi - n}$ obtained in this work (solid red line). The red band corresponds to the $\pm1\sigma$ background-only expectation. We also include 90\% CL exclusion limits from our previous result with a 0.6 kg d exposure \cite{ws2016}, other experiments \cite{otherexps}, and the 90\% CL contours for the WIMP-signal interpretation of the CDMS silicon result \cite{cdms}.}
	\label{fig:excess}
\end{figure}

The developed background model is used to search for light WIMPs ($1$~GeV~$ \lesssim~m_\chi c^2~\lesssim~10$~GeV) below the 6~keV$_\text{ee}$ threshold. The search is carried out on the same 11~kg~d exposure dataset used to build the background model. The profile likelihood ratio method is used to test the background-only hypothesis against the low energy signal hypothesis. In particular, the expected WIMP rate for a given mass is assumed exponentially decaying as a function of energy. Such dependence is enforced onto an otherwise uniform energy-depth signal probability density function. The distribution of observed events is consistent with a background-only scenario down to 200~eV$_\text{ee}$, a sign of thoroughness of the developed background model. Below 200~eV$_\text{ee}$, the fit exhibits a preference for a 17.1$\pm$7.6~events exponential bulk component with a $p$-value of $2.2\times 10^{-4}$. Data points, background model and excess contours are displayed on the left side of Figure \ref{fig:excess}. Several hypothetical explanations for the excess have been explored, particularly focusing on the systematic uncertainties affecting the analysis. No plausible origin for the excess was found among the sources of systematic uncertainty. This finding will be further investigated with the upcoming upgrade of the DAMIC at SNOLAB detector. 

Despite the conspicuous excess observed, we set 90\%~CL upper limits on the WIMP-nucleon cross section. These constitute the most stringent limits from a silicon target experiment for 1~GeV<$m_\chi c^2$~<~9~GeV. The limit plot is reported on the right side of Figure \ref{fig:excess}. 

\section{Future Plans}

The results described in this article constitute further incentive for the ongoing upgrade of DAMIC at SNOLAB. The chance to probe the same background environment with more precise energy and depth reconstruction is offered by the skipper read-out technology. Skipper CCDs non-destructively measure the charge in the same pixel $N$ times to reduce the read-out noise by a factor $\sqrt{N}$. CCDs deploying the skipper technology are capable of resolving single electrons from ionization events. Two 6k$\times$4k DAMIC-M \cite{damicm} and four 1k$\times$6k SENSEI \cite{sensei} skipper CCDs will be operated in the DAMIC at SNOLAB apparatus, which will profit from a few other improvements from the electronics standpoint. With this upgrade, the DAMIC collaboration will be able to assess whether the observed low-energy event excess persists and better investigate its origin.

On a longer term, the DAMIC-M collaboration \cite{damicm} will deploy a kg-scale CCD detector at the Modane underground laboratory (\href{http://www.lsm.in2p3.fr}{LSM}) within the next few years ($\sim 2023$). Striving for the ambitious background goal of 0.1 dru\footnote{1 dru = 1 event kg$^{-1}$d$^{-1}$keV$^{-1}$.} (to contrast with $\sim$10 dru for DAMIC at SNOLAB), it will employ novel electronics developed for the purpose, in addition to implementing the skipper technology. The experiment will feature an energy threshold $\lesssim 10~$eV$_\text{ee}$. These advances broaden the accessible parameter space for dark matter searches with CCDs, and pave the way for a new generation of low-threshold direct detection CCD dark matter experimental projects (\textit{e.g.}, \cite{oscura}).

\clearpage
\section*{Full Authors List: \Coll\ Collaboration}


\scriptsize
\noindent
A.~Aguilar-Arevalo$^1$, 
D.~Amidei$^2$,
I.~Arnquist$^3$,
D.~Baxter$^4$,
G.~Cancelo$^5$,
B.A.~Cervantes Vergara$^1$,
A.E.~Chavarria$^6$,
E.~Darragh-Ford$^4$,
M.L.~Di~Vacri$^3$,
J.C.~D'Olivo$^1$,
J.~Estrada$^5$,
F.~Favela-Perez$^1$,
R.~Ga\"ior$^7$,
Y.~Guardincerri$^{5}$\footnote[2]{Deceased January 2017},  
T.W.~Hossbach$^3$,
B.~Kilminster$^8$,
I.~Lawson$^9$,
S.J.~Lee$^8$,
A.~Letessier-Selvon$^7$,
A.~Matalon$^{4,7}$,
P.~Mitra$^6$,
A.~Piers$^6$,
P.~Privitera$^{4,7}$,
K.~Ramanathan$^4$,
J.~Da~Rocha$^7$,
M.~Settimo$^{10}$,
R.~Smida$^4$,
R.~Thomas$^4$,
J.~Tiffenberg$^5$,
M.~Traina$^7$,
R.~Vilar$^{11}$,
and 
A.L.~Virto$^{11}$ \\
\noindent
$^1$Universidad Nacional Aut{\'o}noma de M{\'e}xico, Mexico City, Mexico.
$^2$Department of Physics, University of Michigan, Ann Arbor, Michigan, United States.
$^3$Pacific Northwest National Laboratory (PNNL), Richland, Washington, United States.
$^4$Kavli Institute for Cosmological Physics and The Enrico Fermi Institute, The University of Chicago, Chicago, Illinois, United States.
$^5$Fermi National Accelerator Laboratory, Batavia, Illinois, United States.
$^6$Center for Experimental Nuclear Physics and Astrophysics, University of Washington, Seattle, Washington, United States.
$^7$Laboratoire de Physique Nucl\'eaire et des Hautes \'Energies (LPNHE), Sorbonne Universit\'e, Universit\'e de Paris, CNRS-IN2P3, Paris, France.
$^8$Universit{\"a}t Z{\"u}rich Physik Institut, Zurich, Switzerland.
$^9$SNOLAB, Lively, Ontario, Canada.
$^{10}$SUBATECH, CNRS-IN2P3, IMT Atlantique, Universit\'e de Nantes, Nantes, France.
$^{11}$Instituto de F\'isica de Cantabria (IFCA), CSIC--Universidad de Cantabria, Santander, Spain.

\end{document}